\documentclass[conference]{IEEEtran}
\IEEEoverridecommandlockouts
\usepackage{cite}
\usepackage{amsmath,amssymb,amsfonts}
\usepackage{algorithmic}
\usepackage{graphicx}
\usepackage{textcomp}
\usepackage{xcolor}
\def\BibTeX{{\rm B\kern-.05em{\sc i\kern-.025em b}\kern-.08em
    T\kern-.1667em\lower.7ex\hbox{E}\kern-.125emX}}

\usepackage[lined, ruled]{algorithm2e}
\usepackage{url}
\usepackage{hyperref}

\newcommand\blfootnote[1]{%
  \begingroup
  \renewcommand\thefootnote{}\footnotetext{#1}%
  \addtocounter{footnote}{-1}%
  \endgroup
}

\newcommand{\METHOD}{\texttt{ActiveTCR}~}
\newcommand{\METHODnospace}{\texttt{ActiveTCR}}

\definecolor{mygray}{gray}{0.7}

\usepackage[normalem]{ulem}

\begin{document}

\title{Active Learning Framework for Cost-Effective TCR-Epitope Binding Affinity Prediction}


\author{
\IEEEauthorblockN{Pengfei Zhang}
\IEEEauthorblockA{
    \textit{Fulton Schools of Engineering (SCAI)}\\
    \textit{Arizona State University}\\
    Tempe, AZ, United States \\
    pzhang84@asu.edu
}
\and
\IEEEauthorblockN{Seojin Bang}
\IEEEauthorblockA{
    \textit{Google DeepMind}\\
    Mountain View, CA, United States \\
    seojinb@google.com
}
\and
\IEEEauthorblockN{Heewook Lee\IEEEauthorrefmark{1}\thanks{\IEEEauthorrefmark{1}Corresponding author.}}
\IEEEauthorblockA{
    \textit{Fulton Schools of Engineering (SCAI)}\\
    \textit{Arizona State University}\\
    Tempe, AZ, United States \\
    heewook.lee@asu.edu
}
}

\maketitle

\begin{abstract}
T cell receptors (TCRs) are critical components of adaptive immune systems, responsible for responding to threats by recognizing epitope sequences presented on host cell surface. Computational prediction of binding affinity between TCRs and epitope sequences using machine/deep learning has attracted intense attention recently. However, its success is hindered by the lack of large collections of annotated TCR-epitope pairs. Annotating their binding affinity requires expensive and time-consuming wet-lab evaluation. To reduce annotation cost, we present \METHODnospace, a framework that incorporates active learning and TCR-epitope binding affinity prediction models. Starting with a small set of labeled training pairs, \METHOD iteratively searches for unlabeled TCR-epitope pairs that are ``worthy'' for annotation. It aims to maximize performance gains while minimizing the cost of annotation. We compared four query strategies with a random sampling baseline and demonstrated that \METHOD reduces annotation costs by approximately 40\%. Furthermore, we showed that providing ground truth labels of TCR-epitope pairs to query strategies can help identify and reduce more than 40\% redundancy among already annotated pairs without compromising model performance, enabling users to train equally powerful prediction models with less training data. Our work is the first systematic investigation of data optimization for TCR-epitope binding affinity prediction.
\end{abstract}


\section{Introduction}
T cell receptors (TCRs) play a pivotal role in adaptive immune systems by recognizing epitope --- a part of antigen--presented on cell surface via  major histocompatibility complex and initiating potential immune responses to safeguard the host~\cite{attaf2015t}. Understanding the binding affinity between TCR and epitope sequences is fundamental for developing immunotherapy strategies, where T cells are engineered/designed and subsequently assessed for their binding results to target epitopes in a wet-lab setting~\cite{schumacher2002t}. However, such assessments are slow and expensive. To overcome these challenges, computational methods for predicting TCR-epitope binding affinity have emerged to streamline the assessment process and minimize expenses.

Many machine learning and deep learning models have been developed to improve the prediction performance of TCR-epitope binding affinity~\cite{springer2021ergo2, montemurro2021nettcr2, cai2022atm, zhang2022pite}. These models typically take two sequences, a TCR and an epitope, as input and predict the binding affinity between them. Despite emergence of these machine learning models, there has been limited attention given to optimizing the underlying data. Most prediction models have primarily focused on exploring the impact of various neural network architectures, neglecting the crucial role of data in achieving optimal performance. 

As databases of annotated TCR-epitope pairs continue to grow~\cite{tickotsky2017mcpas, shugay2018vdjdb, vita2019iedb}, two important research questions arise. First, \emph{how can we reduce the cost of annotating new TCR-epitope pairs in the future?} Reducing the cost of each individual wet-lab experiment is challenging due to the inherent expenses associated with these processes. However, more informed decisions can be made by selectively annotating the ``most important'' or ``most useful'' pairs for prediction models. By intelligently choosing pairs for annotation, we can optimize the allocation of resources, ultimately leading to more efficient use of annotation budgets for TCR-epitope pairs. Second, \emph{how can we optimize the use of those already annotated pairs to train powerful prediction models with less training data without compromising model performance?} This question arises because of the presence of identical or similar TCR-epitope pairs in the data. Some pairs, although not identical, can be semantically similar in the latent feature space and thus contribute minimally to model training. Furthermore, identical pairs, i.e., TCR-epitope pairs with matching sequences, could cause model overfitting and slow down the training process. A noticeable overlap among these pairs has been observed in most of the currently available databases. For example, roughly 49.41\% of TCR-epitope pairs with binding scores greater than zero in VDJdb~\cite{shugay2018vdjdb} are identical to pairs in IEDB~\cite{vita2019iedb}, leading to unnecessary consumption of computational resources. These similar pairs are considered less important or less beneficial to the model. Therefore, our objective is to identify and reduce the redundancy inherent among these already annotated TCR-epitope pairs. By doing so, users can train equally powerful prediction models with less training data.

In this study, we present \METHODnospace, a novel framework designed to address research questions of reducing annotation cost for future unlabeled pairs and reducing data redundancy among already labeled data in the context of TCR-epitope binding affinity prediction. \METHOD employs an active learning approach, where a prediction model is initially trained on a small subset of the training data then the model continuously selects the most informative samples from the unlabeled TCR-epitope pool to be annotated and retrain the prediction model until satisfactory performance is achieved or no more samples are available. 

The crux of our framework lies in querying the ``most important'' pairs for the TCR-epitope binding affinity prediction model. We utilized entropy as a heuristic measure of ``importance'' of each pair concerning prediction models. To investigate the effectiveness of \METHODnospace, we explored five distinct query strategies, including a random sampling baseline, three variants of entropy sampling, and a misclassification sampling strategy for reducing redundancy. We evaluated \METHOD in two scenarios. Our experimental results demonstrated that \METHOD effectively halved the annotation cost for future unlabeled TCR-epitope pairs and reduced over 40\% data redundancy among those already annotated pairs. \METHOD provides a promising solution to the challenges of reducing the annotation cost for future data and reducing redundancy among already annotated data for TCR-epitope binding affinity prediction models. \blfootnote{Code and models are publicly available at \url{https://github.com/Lee-CBG/ActiveTCR}. 
} 

\section{Related Work}
\subsection{Computational Approaches for Binding Affinity Prediction}
In order to predict binding affinity between TCR and epitope sequences, researchers have spent large efforts in employing machine learning and deep learning techniques. Two primary approaches explored to improve prediction performance are 1) designing the neural network architecture of prediction models and 2) developing amino acid embedding models for TCR and epitope sequences. Early models focused on neural network structures of prediction models while using a simple embedding matrix, BLOSUM62~\cite{henikoff1992blosum}, to map amino acid residues to continuous numeric vector representations. For example, NetTCR~\cite{montemurro2021nettcr2} used a series of convolutional layers to learn features of input TCR and epitope sequences whereas ERGO2~\cite{springer2021ergo2} proposed an LSTM-based approach to learn sequential information of amino acid sequences. Later, ATM-TCR~\cite{cai2022atm} utilized multi-head attention modules to learn contextualized features of amino acids. While these models achieved fair prediction performance on known epitopes (AUCs of 72.0--77.3\%), they performed poorly in correctly identifying binding TCRs for novel (unseen) epitopes that were not observed during training (AUCs of 47.0--54.2\%).

To develop models that can better generalize to unseen epitopes, researchers incorporated binding affinity prediction models with advanced amino acid embedding techniques. Several such models have been proposed to learn representations from a large corpus of TCR sequences, which can be used as feature extractors for the input sequences of TCR-epitope binding affinity prediction models. TCR-BERT~\cite{wu2021tcrbert} learned amino acid embeddings using a masked language model, with its architecture inspired by the well-known language model BERT~\cite{devlin2018bert}. catELMo~\cite{Zhang_2023}, a model inspired by ELMo~\cite{peters2018contextualized}, learned amino acid embeddings by predicting the next token based on its previous tokens processed by a stack of bi-directional LSTM layers. Embeddings from catELMo led to significant performance gains for unseen epitope sequences~\cite{zhang2022pite, Zhang_2023}. It was also demonstrated that a prediction model using catELMo embeddings with only 10\% of the training data outperformed one using BLOSUM62 embedding with 100\% of data, indicating potential in reducing the annotation cost of TCR-epitope pairs.

\subsection{Active Learning}
Active learning is a machine learning algorithm that interactively requests annotations of new unlabeled training samples. Unlike passive learning, where a model is trained on a fixed set of labeled data, active learning incorporates human annotators (oracles) to iteratively label the most informative samples. Various query strategies have been developed to identify such informative samples for annotation and inclusion in the training set. Uncertainty sampling~\cite{seung1992query} selects the samples that a model is most uncertain about. The intuition is that the samples that confuse the model the most will benefit its learning the most. A common example for selecting high uncertainty samples is entropy-based strategy~\cite{holub2008entropy, ozdemir2018active}. It selects samples for which the predicted probability distribution over all labels is uniformly spread out. Diversity sampling measures the prediction diversity of unannotated samples, using this diversity as an indicator of informativeness~\cite{settles2007multiple, Ducoffe2017ActiveLS, tsymbalov2018dropout}. 

By updating the training set, active learning can improve model performance while reducing annotation costs. This approach is particularly useful when labeled data is costly and scarce but unlabeled data is abundant. Active learning has proven effective in a variety of fields, including natural language processing~\cite{siddhant-lipton-2018-deep}, computer vision~\cite{haussmann2020scalable,yuan2021multiple}, and medical image analysis~\cite{budd2021survey}. It has also been applied to improve molecular-level optimizations such as prediction of protein-protein interaction~\cite{mohamed2010active} and target-drug interaction~\cite{murphy2011active, eisenstein2020active}. To the best of our knowledge, no prior work has explored active learning approaches in the context of TCR-epitope binding affinity prediction.

\section{Data}
This section describes how we prepared TCR-epitope pairs and divided them into training and testing sets. As the third complementary-determining region (CDR3) of the TCR \(\beta\) chain is the most critical component that interacts with epitope sequences~\cite{la2018understanding}, we made use of the CDR3 of TCR $\beta$ chain and referred to it as TCR unless otherwise specified.

\subsection{TCR-epitope Pairs}
\subsubsection{Positive Pairs}
We sourced TCR-epitope pairs of human species from three publicly available databases: VDJdb~\cite{shugay2018vdjdb}, McPAS~\cite{tickotsky2017mcpas}, and IEDB~\cite{vita2019iedb}. These pairs are clinically known to bind to each other and were used as our positive data. We followed the same pre-processing procedure as ATM-TCR~\cite{cai2022atm}. The data was filtered to only include pairs of human MHC class I epitopes and TCR$\beta$ sequences. We kept pairs with linear epitope sequences and discarded pairs containing wildcards such as * or X in sequences. For VDJdb~\cite{shugay2018vdjdb}, we only included pairs with a confidence binding score greater than zero. We also eliminated any duplicated TCR-epitope pairs, resulting in 150,008 unique TCR-epitope pairs, with 982 unique epitopes and 140,675 unique TCRs.

\subsubsection{Negative Pairs}
Given the scarcity of clinically confirmed negative TCR-epitope pairs~\cite{montemurro2021nettcr2}, generating negative pairs is a standard practice in the field of TCR-epitope binding prediction~\cite{montemurro2021nettcr2, cai2022atm, zhang2022pite, Zhang_2023}. In order to synthesize negative pairs, we adopted the strategy of pairing existing epitopes with newly sampled TCRs from repertoires, an approach that has been substantiated in prior studies~\cite{montemurro2021nettcr2, Zhang_2023, zhang2022pite}. The rationale behind this approach was to simulate training pairs where TCRs that are prevalent in healthy individuals do not bind with disease epitopes, thereby providing a comprehensive contrast to the positive pairs in our machine learning model. We first randomly sampled TCRs from 20 million TCR sequences of healthy control repertoires of ImmunoSEQ~\cite{nolan2020immuneSeq}. We then replaced the original TCRs of the positive TCR-epitope pairs with TCRs of the healthy repertoires. This resulted in 150,008 unique negative TCR-epitope pairs.

\subsection{Training and Testing Set Splits}
It is of interest to measure binding affinity prediction performance on epitopes and TCRs that have never been observed before. A random split was not suitable for measuring generalization performance, particularly for unseen epitopes. Given that 99.97\% of epitopes occur multiple times in our dataset, it is highly likely that an epitope would be present in both the training and testing sets. Although 96.8\% of TCRs in our dataset are unique, the same issue may still occur for TCR sequences. To accurately assess the prediction performance on the novel (unseen) TCRs and epitopes, we used the two data partition approaches in ATM-TCR~\cite{cai2022atm}. The epitope split shared no common epitopes between training and testing sets, allowing us to evaluate the model's generalizability on novel epitope sequences. Similarly, the TCR split shared no common TCRs between training and testing sets, enabling us to evaluate the model's generalizability on novel TCR sequences. For both splits, we used 80\% of the entire data as training data and the remaining 20\% as testing data.

\section{Methods}
\subsection{\METHOD} 

\METHOD is an active learning framework designed for rapid and cost-effective development of TCR-epitope binding affinity prediction models. The crux of \METHOD lies in its ability to interactively query a large size of pool to iteratively update the training set \(D\) by adding the most informative TCR-epitope pairs, thereby learning improved prediction models \(M\). \METHOD has two use cases in the context of TCR-epitope binding affinity prediction: 1) reducing annotation costs of unlabeled TCR-epitope pairs and 2) reducing redundancy among already labeled pairs. 

\begin{figure*}[htbp]
  \centering 
  \includegraphics[width=1.5\columnwidth]{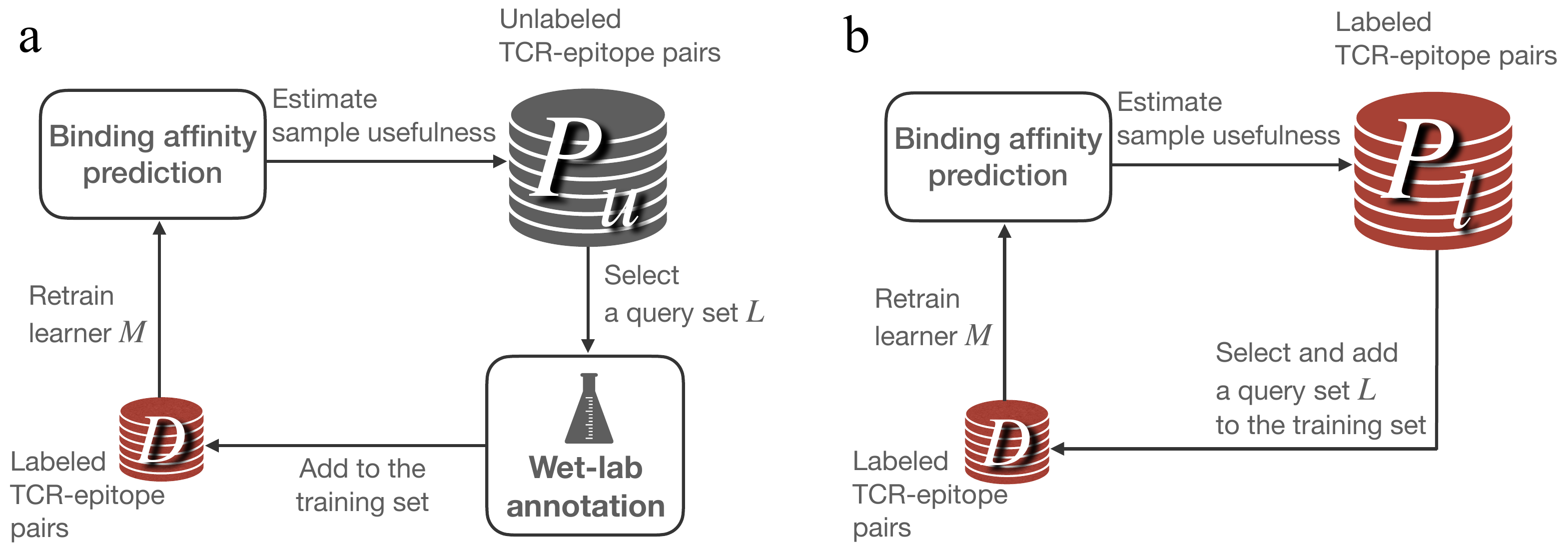} 
  \caption{\METHOD framework for TCR-epitope binding affinity prediction. It can be used \textbf{a)} to reduce annotation cost for unlabeled TCR-epitope pairs, and \textbf{b)} to reduce data redundancy among labeled TCR-epitope pairs. It interactively queries the most informative TCR-epitope pairs to iteratively construct a training set and to update a binding affinity prediction model.}
  \label{fig:method} 
\end{figure*}

The goal of the first use case is to maximize prediction performance gains while minimizing the associated annotation costs. As shown in Fig.~\ref{fig:method}a, \METHOD iteratively queries the most informative TCR-epitope pairs to be annotated by wet-lab efforts. The process begins by training a TCR-epitope binding affinity prediction model \(M\), referred to as a learner, on an initial training set of labeled TCR-epitope pairs \(D\). The learner \(M\) then interactively queries unlabeled TCR-epitope pairs from the pool \(P_u\). In order to do so, it predicts binding affinities of the unlabeled pairs and selects the most informative pairs by a query strategy \(Q\). Wet-lab annotators then provide the ground truth binding labels for the queried TCR-epitope pairs \(L\). The newly labeled pairs are added to the training set \(D\), and the model \(M\) is retrained on the updated training set \(D\). Through this feedback loop, \METHOD continually updates model \(M\) until satisfactory prediction performance is achieved. Detailed procedure is provided in Alg.~\ref{alg_activeTCR}.

The goal of the second use case is to develop a data efficient model that can achieve the same or better prediction performance with a less number of TCR-epitope training pairs. As shown in Fig.~\ref{fig:method}b, \METHOD iteratively queries more labeled pairs \(L\) and adds them to the training set \(D\) to improve the prediction model \(M\). Redundant pairs may include identical data entries found in different databases or semantically similar entries that do not significantly contribute to the model performance. Since the ground truth label is accessible in the pool \(P_l\), we can leverage it for querying informative pairs and discard redundant pairs. The wet-lab annotators are no longer required in this setting. Detailed procedure is provided in Alg.~\ref{alg_activeTCR}.

\begin{algorithm}
    \SetAlgoLined
    \SetKwInOut{Inputs}{Inputs}
    \SetKwInOut{Outputs}{Outputs}
    \Inputs{$D = {(X_{tcr}, X_{epi}, y)}$: Initial training set \\
            $P_u = {(X_{tcr}, X_{epi})}$: Unlabeled TCR-epitope pool \tcp{only for use case \textsf{a}} \\
            $P_l = {(X_{tcr}, X_{epi}, y)}$: Labeled TCR-epitope pool \tcp{only for use case \textsf{b}} \\
            $Q$: Query strategy \\
            $k$: Query size per iteration
             }
    \Outputs{$M_t$: Optimized prediction model at iteration $t$ \\ 
             $L_t$: Queried TCR-epitope pairs at iteration $t$
             }
    \SetKw{Initialize}{Initialize}
    \SetKw{Update}{update}
    \SetKw{Select}{select}
    \SetKw{Repeat}{Repeat}
    \SetKw{Compute}{compute}
    \SetKw{Return}{return}
    \Initialize{$L_0 \gets \varnothing $, $t \gets 0$} 
    
     \While{stopping criterion is not met}{

        Train TCR-epitope binding affinity prediction model $M_t$ on $D$\;
        
        \uIf{use case \textsf{a}: reduce annotation cost}{
                 $L_{t+1} \gets Q(M_t, P_u, k)$ \tcp{select $k$ most informative samples from $P_u$} 
                 
                 Obtain wet-lab annotation of $L_{t+1}$\;
                 
                 $P_u \gets P_u \setminus L_{t+1}$\; 
        
              }

              \uElseIf{use case \textsf{b}: decrease computational cost}{
                $L_{t+1} \gets Q(M_t, P_l, k)$ \tcp{select $k$ most informative samples from $P_l$}
                
                $P_l \gets P_l \setminus L_{t+1}$\;
                }
        
        $D \gets D \cup L_{t+1}$
              
        $t \gets t + 1$
        
    }
    \caption{\textsc{\METHOD}}
    \label{alg_activeTCR}
\end{algorithm}

\subsection{Query Strategies} 
\paragraph{Global and local entropy-based sampling}
Entropy-based sampling~\cite{lewis1995sequential} queries TCR-epitope pairs that the model is least certain about their binding status. Such pairs are typically located near the classification boundary, making them more informative about shape of the boundary than pairs located farther from it. Entropy~\cite{shannon1948mathematical} is an intuitive metric for assessing the uncertainty of machine learning model predictions. A low entropy score of a sample indicates that the model is more confident in its prediction result, while a high entropy score suggests less confidence. The entropy score of a model $M$'s prediction for a pair of TCR and epitope ($t_i, e_i$) is defined as follows:


\begin{multline}
Q_{EN}(t_i, e_i; M) = H(\hat{y}_i) \\
= - \left(\hat{y}_i\log_2 \hat{y}_i + (1-\hat{y}_i)\log_2 (1-\hat{y}_i) \right)
\end{multline}

\noindent where $\hat{y}_i$ is model \(M\)'s prediction output for the 
TCR-epitope pair ($t_i, e_i$). At each iteration, prediction model \(M\) predicts binding affinity \(\hat{y}\) for pairs in \(P_u\) (or \(P_l\)) and selects those with the highest entropy scores to construct new training sets \(D\). We used two entropy-based query strategies: \textit{global entropy sampling} and \textit{local entropy sampling}. Global entropy sampling queries pairs from the entire pooled pairs \(P_u\) (or \(P_l\)) at each iteration, while local entropy sampling queries pairs from a random subset of the pool, $P_{us}\in P_u$ (or $P_{ls}\in P_l$). In our experiment, we set the size of \(P_{us}\) (or \(P_{ls}\)) to be twice the size of the query pairs. For example, if 1,000 query pairs are selected at each iteration, the size of \(P_{us}\) (or \(P_{ls}\)) would be 2,000. Global entropy sampling is more reliable than local entropy sampling as it calculates entropy scores across all unlabeled pairs. Meanwhile, local entropy sampling is faster as it only computes entropy for a subset of pooled pairs. 

\paragraph{Query-by-dropout-committee sampling}
Query-by-committee~\cite{seung1992query} selects samples based on the level of disagreement among multiple prediction models. Each model, known as a committee member, learns its own decision boundary and predicts the binding affinity of a TCR-epitope pair. If the committee members largely disagree on their predictions, the pair is considered uncertain and added to training sets \(D\). This strategy tends to be slower and more computationally demanding because it requires training multiple prediction models to serve as committee members. To address these limitations, researchers have used dropout during inference time to measure committee disagreement in active learning frameworks \cite{Ducoffe2017ActiveLS, tsymbalov2018dropout}. This technique enables dropout layers during the prediction phase, introducing variation to the model's prediction outputs by randomly dropping out neurons. As a result, the same model can produce different binding affinity scores for a pair of TCR and epitope without the need of training multiple committee models. This approach, referred to as dropout committee, is commonly used to measure the uncertainty of a neural network model with reduced computational burden~\cite{gal2016mcdropout}. 

We quantified the disagreement score for a TCR-epitope pair by a sum of Kullback-Leibler divergence~\cite{kullback1951information} between each test time prediction and their average:


\begin{multline}
Q_{C}(t_i, e_i; M) = \sum_{j} D_{KL}\left(\widehat{Y}_{i,j}\ ||\ \overline{Y}_{i}\right) \\
= \sum_{j}\sum_{k \in \mathbf{0,1}} \hat{y}_{i,j,k} \log_2\left(\frac{\hat{y}_{i, j,k}}{\overline{y}_{i,k}}\right)
\end{multline}

where $k$ is possible prediction outcomes ($0$ for non-binding and $1$ for binding in our case), $\hat{y}_{i,j,k}$ is the binding affinity prediction score of $(t_i, e_i)$ made by $j$-th committee member, and $\overline{y}_{i,k}$ is an average of all committee members' prediction scores. In our experiment, we set the number of dropout committees as 10.

At each iteration, we applied \textit{dropout committee sampling} to a randomly selected subset of pooled pairs \(P_{us}\) (or \(P_{ls}\)) for computational efficiency. To investigate whether this query strategy can enhance the performance of entropy-based queries, we added it to the local entropy strategy, defining the measurement as follows. 
\begin{equation}
Q_{EN,C}(t_i, e_i; M) = \omega Q_{EN}(t_i, e_i; M)  + (1-\omega) Q_{C}(t_i, e_i; M) 
\label{equation:local_n_dropout}
\end{equation}
where $\omega$ is a weight between local entropy sampling and dropout committee sampling. For simplicity, we assumed the same importance for those two samplings and assigned equal weights to them in our experiments.

\paragraph{Misclassification sampling}
We designed a simple yet effective query strategy called \textit{misclassification sampling}. It selects samples from the labeled TCR-epitope pool $P_l$ that are misclassified (both incorrectly labeled as positive and incorrectly labeled as negative) by the model $M$ with a significant difference between the true label and the predicted label. The intuition is that samples that are misclassified by a large margin can potentially improve model performance by providing more information about the model's weaknesses. The misclassification sampling score of a TCR-epitope pair $(t_i, e_i)$ is defined as follows:
\begin{equation}
MC(t_i, e_i, y_i; M) = \left| y_i - \hat{y}_i \right| \text{ for } i \in P_l
\end{equation}
where $y_i$ is the ground truth binary label and $\hat{y}_i=M(t_i, e_i)$ is the binding affinity prediction score between 0 and 1. Note that this approach is only applicable to the use case of reducing redundancy among annotated TCR-epitope pairs as it requires ground truth labels for them in $P_l$ to determine misclassification.

\subsection{Binding Affinity Prediction Model}
Our learner \(M\) is a TCR-epitope binding affinity prediction model that predicts the likelihood of binding between a TCR and an epitope. It takes two sequence inputs (a TCR and an epitope) and outputs a probability that the TCR will bind to the epitope. In our active learning framework, users have the flexibilities to select the binding affinity prediction model that best suits their needs, for instance, ATM-TCR~\cite{cai2022atm}, ERGO2~\cite{springer2021ergo2}, NetTCR~\cite{montemurro2021nettcr2}, and PiTE~\cite{zhang2022pite}.

We used the 3-linear-layered model from catELMo~\cite{Zhang_2023} as our prediction model \(M\) because it performed similarly to the state-of-the-art performance of PiTE while being faster and more lightweight~\cite{zhang2022pite}. It is composed of two main steps: amino acid embedding and binding affinity prediction. In the first step, the TCR and epitope sequences were represented as numeric vectors of size 1,024 using catELMo, a state-of-the-art amino acid embedding model. In the second step, a simple prediction model with three linearly connected layers was trained to predict the binding affinity between the embedded TCR and epitope sequences. The embedded vectors were used as input for the model. Each was first processed through a linear layer with 2,048 neurons followed by a Sigmoid Linear Units (SiLU) activation function~\cite{elfwing2018sigmoid}, batch normalization~\cite{ioffe2015batch}, and 0.3 rate dropout~\cite{srivastava2014dropout}. The two processed sequences were then concatenated and fed to a linear layer with 1,024 neurons, followed by a SiLU activation function, batch normalization, and 0.3 rate dropout. Finally, the last linear layer with a neuron followed by a sigmoid activation function produced a binding affinity score between 0 and 1. Binary cross-entropy loss and Adam optimizer~\cite{kingma2014adam} were used to train the model. The batch size was 32 and the learning rate was 0.001. The training continued for 200 epochs or was early terminated if the validation loss did not decrease for 30 consecutive epochs.

\section{Results on Real Data} 
\subsection{Study Design} 

In this section, we experimentally assess how \METHOD contributes to the annotation cost reduction for unlabeled TCR-epitope pairs and to the data redundancy reduction among labeled pairs.

We randomly selected 10\% (about 24,006 pairs) of the training TCR-epitope pairs as the initial training data \(D\). The remaining 90\% (about 216,054 pairs) of the training set was treated as pool data (\(P_u\) or \(P_l\)). Our learner model \(M\) predicted the binding affinities \(\hat{y}\) and identified the most informative pairs based on the different query strategies. At each iteration, 24,006 additional pairs \(L\) were queried from pool data. For the use case of reducing annotation costs of unlabeled data, we removed the binding labels of TCR-epitope pairs in \(P_u\) to simulate the presence of future unlabeled pairs. The ground truth labels were provided to the model only after it made queries \(L\), and served as the wet-lab annotation. 
For the use case of reducing data redundancy, we allowed the query strategies to leverage ground truth label information to select pairs from \(P_l\). \METHOD continuously expanded the size of dataset \(D\) by adding the queried samples \(L\). The prediction model \(M\) was interactively retrained until all pairs in the pool were exhausted or satisfactory results were achieved. 
The testing set was held out to assess the performance of the model \(M\) at each iteration. We reported AUCs of prediction models using different query strategies on the testing set. Each query strategy had 10 independent runs with a seed value of 42 unless otherwise specified. 

\begin{figure}[h]
  \centering 
  \includegraphics[width=.9\columnwidth]{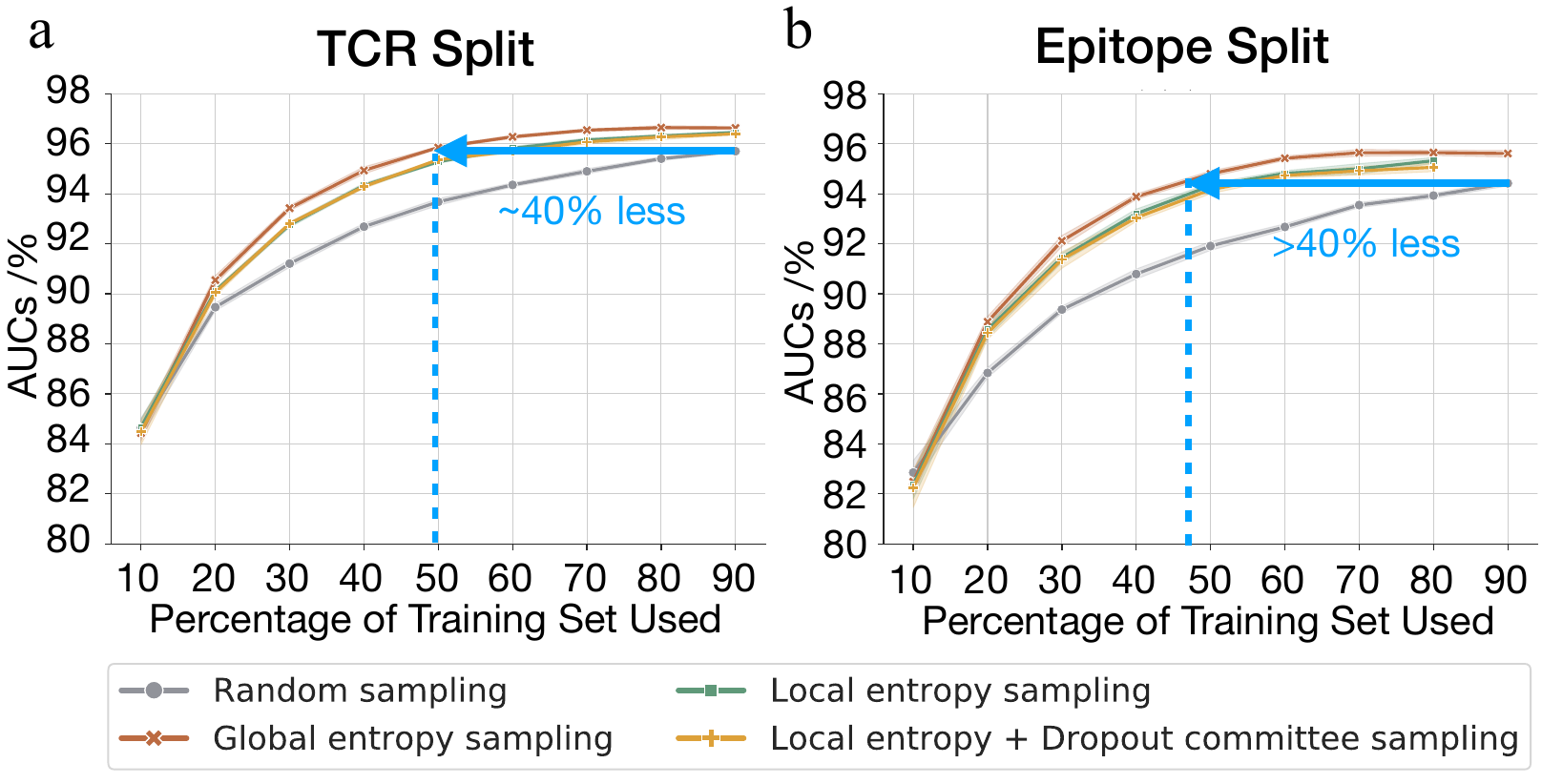} 
  \caption{Performance of \METHOD in reducing annotation costs, measured on \textbf{a)} TCR split and \textbf{b)} epitope split. Average (solid line) and standard error (band) AUC of 10 independent runs for each query strategy are reported. \METHOD using global entropy sampling reduced approximately 40\% annotation costs compared to the random sampling baseline. Two of the query strategies required sub-sampling the unlabeled pool, which was earlier stopped at iteration 8 (80\% training set) in epitope split due to insufficient TCR-epitope pairs for sub-sampling.
}
  \label{fig:main_reduce_annotation_cost} 
\end{figure}

\subsection{Reduce Annotation Cost by Interactively Annotating Unlabeled Samples} 

\METHOD significantly reduced annotation costs for future unlabeled TCR-epitope pairs compared to the random sampling baseline. We simulated a realistic scenario in which not all labels are available at the beginning. We assumed that only 10\% samples are labeled, which is our initial set. We then iteratively added back the ``labels" of each additional 10\% training samples (to mimic the wet lab procedure of annotating additional TCR-epitope pairs) by \METHODnospace. The key question we addressed in this experiment is whether \METHODnospace, as an active learning framework, can select the most informative TCR-epitope pairs to be annotated, such that it achieves a higher performance gain compared to a random sampling strategy.

Among the strategies, global entropy sampling reduced at least 40\% annotation cost in both TCR and epitope splits (Fig.~\ref{fig:main_reduce_annotation_cost}). Global entropy sampling, while being more computationally intensive than local entropy sampling, was found to be more effective in reducing annotation cost and improving the model performance. This suggested that samples queried from global entropy were more informative and therefore contributed more to the model's learning than those queried using local entropy sampling. Local entropy with dropout committee sampling did not appear to offer additional benefits over local entropy sampling in both TCR and epitope splits, indicating that using a dropout layer to introduce uncertainty does not necessarily improve model learning. A possible explanation is that the weight (\(w\) in Equation~\ref{equation:local_n_dropout}) we assigned between them is suboptimal. Misclassification sampling was excluded from the comparison as it required ground truth labels for pairs.

As the model queried new samples with their binding affinity ground truth hidden at each iteration, we did not have control over the ratio of queried positive and negative pairs at each iteration. To understand the preference of each query strategy over positive and negative pairs, we visualized the amount of queried positive and negative pairs in \(L\) for each iteration in both TCR and epitope splits (Fig.~\ref{fig:query_distributions_all_methods}). We found that global entropy sampling, the best-performing strategy, had a relatively uneven distribution of positive and negative pairs at each iteration but maintained a relatively balanced positive-negative ratio for cumulative pairs \(D\) (Fig.~\ref{fig:query_distribution_individual_epitopes}). 

\begin{figure}[h]
  \centering 
  \includegraphics[width=.9\columnwidth]{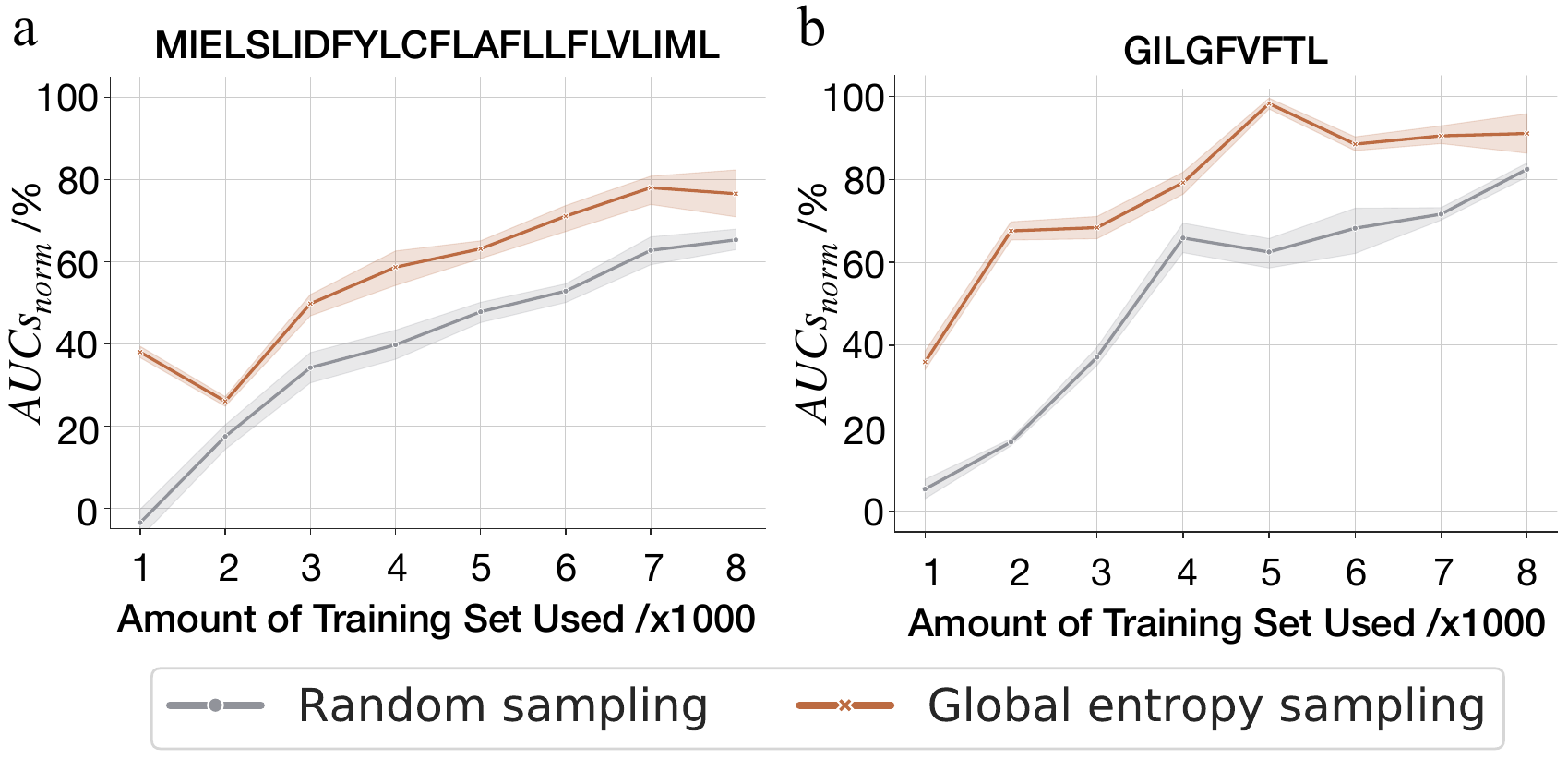} 
  \caption{Epitope-specific performance of \METHOD in reducing annotation costs. The performance is measured on epitope split for two individual epitopes \textbf{a)}  \texttt{MIELSLIDFYLCFLAFLLFLVLIML} and \textbf{b)} \texttt{GILGFVFTL}. The average (solid line) and standard error (band) of \(AUC_{norm}\) from 5 independent runs for each query strategy are reported. 
  \(AUC_{none}\) for \texttt{MIELSLIDFYLCFLAFLLFLVLIML} and \texttt{GILGFVFTL} are 81.80\% and 92.98\% when fine-tuning \(M_0\) on zero pairs, respectively. \(AUC_{all}\) for \texttt{MIELSLIDFYLCFLAFLLFLVLIML} and \texttt{GILGFVFTL} are 89.04\% and 95.47\% when fine-tuning \(M_0\) on the entire new unlabeled pool, respectively. \METHOD consistently outperforms the random sampling baseline with however many TCR-epitope pairs queried and annotated. 
  }
  \label{fig:main_individual_epi} 
\end{figure}

We also showed that \METHOD consistently improved prediction performance for individual novel epitopes by fine-tuning the initial prediction model \(M_0\) on queried pairs. Such investigation is particularly beneficial when the goal is to optimize the prediction performance of novel or rarely observed target epitopes associated with specific diseases. We emulated a situation where a single laboratory, equipped with an initial prediction model trained on existing TCR-epitope pairs, aims to explore a novel target epitope. This setup aligns with the laboratory's research interests as it represents a common real-world scenario where labs are often interested in studying novel or disease-specific epitopes. We assumed that the initial prediction model was trained on an existing database of multiple epitopes and TCRs, and that the target epitope was novel to the model. To ensure the target epitope has never been seen by the model, we selected the epitope from the testing set of epitope split. The initial prediction model \(M_0\) was trained on 10\% of randomly selected pairs from the epitope split training set. We did not utilize the remaining pairs from the training set as we assumed only a limited number of training pairs are available. We then used 90\% of the randomly selected TCR and the target epitope pairs from the epitope split testing set as the unlabeled pool, and the remaining 10\% as testing pairs. 

We fine-tuned the initial model \(M_0\) on different sizes of query sets ($1, \cdots, 8K$). A 10x smaller learning rate of 0.0001 was used to prevent model overfitting. We compared the best-performing global entropy sampling query with a random sampling baseline and quantified the performance improvements. Fig.~\ref{fig:main_individual_epi} shows results for the two most abundant epitopes in the testing set, \texttt{MIELSLIDFYLCFLAFLLFLVLIML} and \texttt{GILGFVFTL}. We reported the normalized AUC (\(AUC_{norm}\)) and defined it as follows:

\begin{equation}
AUC_{norm} = (AUC - AUC_{none}) / (AUC_{all} - AUC_{none})
\end{equation}

where \(AUC_{none}\) is the AUC of the initial model that fine-tuned on zero queried pairs, and \(AUC_{all}\) is the AUC of the fine-tuned model on the entire unlabeled pool pairs of an epitope. Our results indicated that \METHOD with global entropy sampling consistently outperforms the random sampling baseline in terms of prediction performance for individual epitopes, regardless of the number of TCR-epitope pairs queried and annotated. This experimental setup illustrated how \METHODnospace's query aligns with a laboratory's interests and capabilities, assisting in the selection of the most ``promising'' TCR-epitope pairs for testing, thereby reducing the experimental workload while maximizing the potential discovery of useful knowledge.

\subsection{Identify and Reduce Redundancy among Labeled Samples}

\begin{figure}[htbp]
  \centering 
  \includegraphics[width=.9\columnwidth]{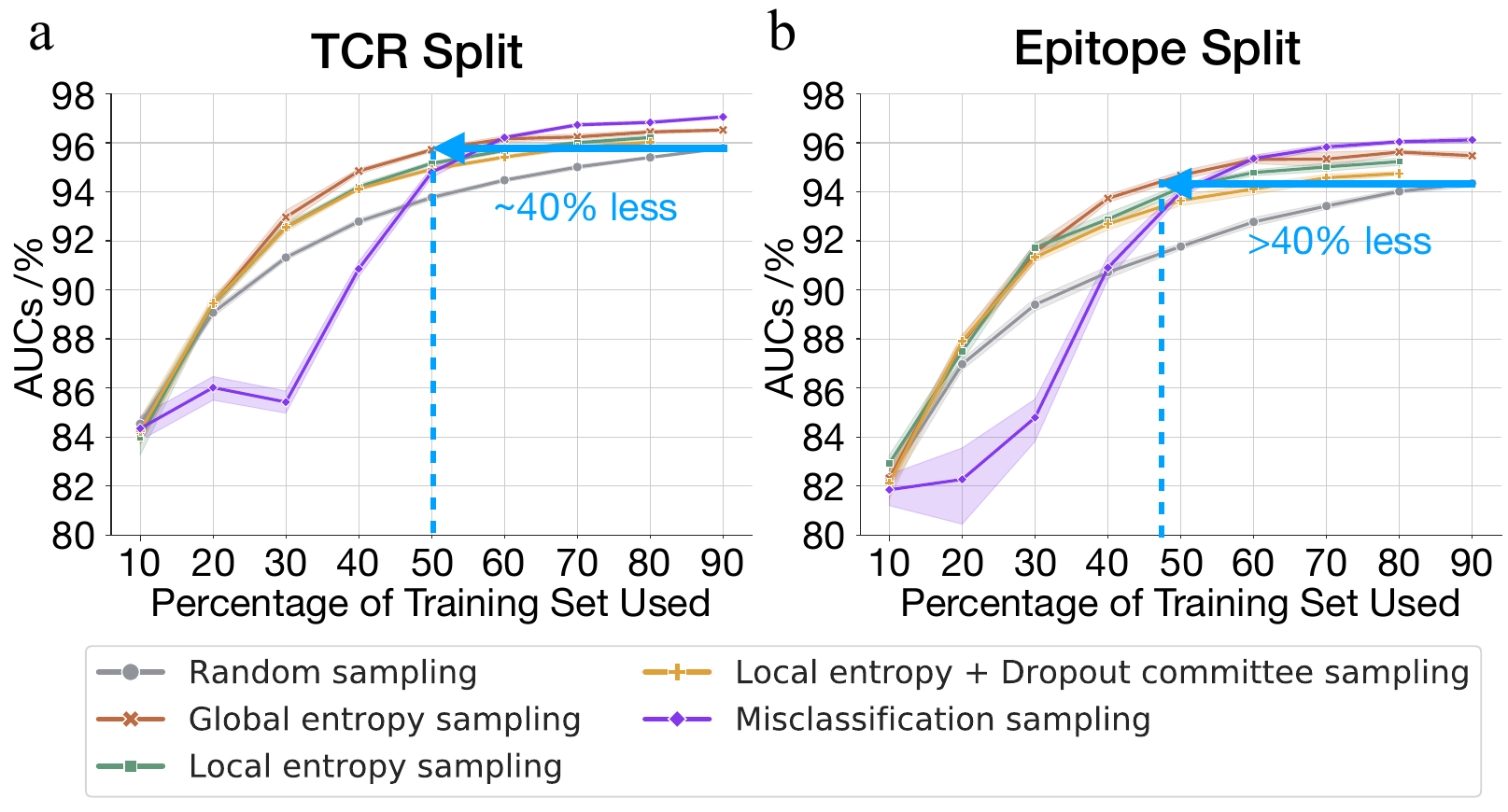} 
  \caption{\METHOD performance comparison in reducing computational cost by reducing redundancy for TCR-epitope pairs using five different query strategies for \textbf{a)} TCR split and \textbf{b)} epitope split. Average (solid line) and standard error (band) AUC of 10 independent runs for each query strategy are reported. \METHOD using global entropy sampling reduced approximately 40\% redundancy among labeled data compared to the random sampling baseline. Two of the query strategies required sub-sampling the unlabeled pool, which was earlier stopped at iteration 8 (80\% training set) in both TCR and epitope split due to insufficient TCR-epitope pairs for sub-sampling. 
  }
  \label{fig:main_reduce_redundancy} 
\end{figure}

\METHOD identified and removed at least 40\% of training data as redundancy (Fig.~\ref{fig:main_reduce_redundancy}) while matching the performance of passive learners with random sampling. This significant reduction directly translates into computational savings for future model training tasks. In this experiment, we queried positive and negative pairs separately. This gave us control over the distribution of queried positive and negative pairs. We queried 5\% (12,003 pairs) from each at each iteration. \METHOD with global entropy sampling with only 50\% of the training data achieved similar performance to random sampling with 90\% of training data, indicating that more than 40\% of the training data were not necessarily contributing to the model. 

We noticed that, with misclassification sampling, \METHOD struggled to achieve high AUC prediction scores with a larger number of queried training pairs at first. This may be because the model was being fed with ``difficult'' samples, making it challenging to generalize well on the testing set. However, by the third iteration, the performance score of the method started to improve and even surpassed that of global entropy sampling in the final iterations. 

We also observed that global entropy consistently outperformed local entropy in reducing data redundancy, suggesting that samples queried from global entropy are less redundant and could improve the model’s performance more. Despite its relatively lower performance compared to other query strategies, local entropy with the dropout committee sampling was still able to reduce approximately 30\% of redundant data in the epitope split and 20\% in the TCR split. We speculated that the reason for its suboptimal performance is the challenge of determining the weights of local entropy and the dropout committee sampling method. Nonetheless, this approach can still provide a significant reduction in annotation costs. Overall, we demonstrated that \METHOD is effective for reducing redundancy among those TCR-epitope pairs with annotated binding results. It can significantly reduce the amount of training data required to match comparable performance to passive learning.

It should be noted that the goal of this section is to 1) demonstrate that there are many redundant TCR-epitope pairs that do not necessarily add to performance gains of the prediction model and to 2) identify a compact, highly informative subset of the data (which we refer to as the ``primal dataset''). When new TCR-epitope pairs are added to a dataset, the ``primal data'' computed prior to the addition of new pairs can be used as a starting point and querying can be done only from the new pairs. This avoids the need to retrain prediction models from scratch using the entire data including the newly added pairs. We conducted an additional experiment and observed no significant performance differences between utilizing an identified primal dataset and running from scratch. In this experiment, we used 60\% of the training data (referred to as $A$) as initially available TCR-epitope pairs with labels. Running \METHOD on this dataset ($A$), a subset was identified as primal data ($A_{Primal}$) after removing $A_{Redundant}$, where $A=A_{Primal} \cup A_{Redundant}$,  and $A_{Primal} \cap A_{Redundant} = \emptyset$.
Then, we randomly selected a set of additional TCR-epitope pairs with labels (referred to as $B$). The total number of pairs in $B$ is one-third of the number of pairs in $A$. Note that $A$ and $B$ are disjoint. With a larger number of labeled pairs, in order to newly define the primal data ($(A \cup B)_{Primal}$), one can either rerun \METHOD on the entirety of $A \cup B$, or utilize the previously defined primal data ($A_{Primal}$) as the starting point and query only from the newly added dataset (B). Utilizing the previously defined primal data only requires iterative model training sweeping through the additional data $B$. However, training from scratch requires more iterations of model training. This reduction in training iterations grows with each occasion of obtaining new labeled data. 
%
In the TCR split, the former approach yielded an average AUC score of 96.09\%, while the latter resulted in a score of 95.35\%. Similarly, in the epitope split, the former method achieved an AUC of 95.42\%, and the latter, 94.41\%. These similar performances indicate \METHOD allows users to expand upon pre-existing datasets without the need to rerun \METHOD from the beginning.

\section{Discussion} 
We proposed \METHODnospace, an active learning framework for cost-effective TCR-epitope binding affinity prediction. We investigated five query strategies and demonstrated the advantages of \METHOD in two practical scenarios, reducing annotation costs of unlabeled pairs and decreasing computational cost by reducing redundancy among annotated TCR-epitope pairs. 

Despite the significant reduction in annotation cost and data redundancy achieved by \METHODnospace, retraining the prediction model \(M\) at each iteration is relatively slow. To address this, we investigated an alternative training strategy: fine-tuning of the previous prediction model on newly queried samples \(L\), instead of training a new model on updated training sets \(D\). We found that the fine-tuning method converges much faster but severe overfitting problems were also observed for all query strategies. 

\begin{figure}[h]
  \centering 
  \includegraphics[width=\columnwidth]{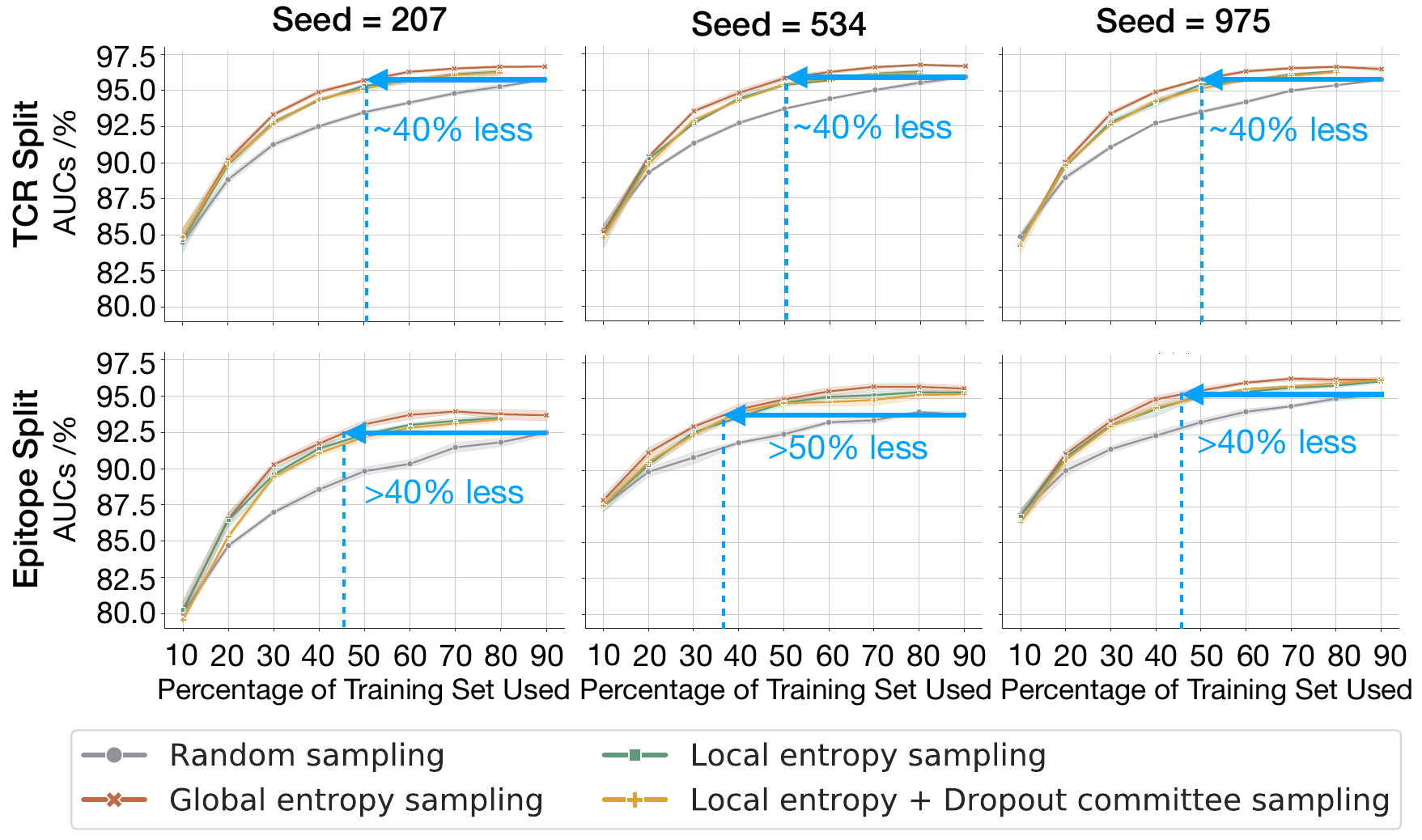} 
  \caption{Performance of \METHOD in reducing annotation costs for different initial training sets. Each row represents TCR split (top) and epitope split (bottom). Each column represents different random seeds to choose the initial training set. Average (solid line) and standard error (band) AUC of 5 independent runs for each query strategy are reported. The amount of annotation cost saved using \METHOD is represented by blue arrows and numbers.
}
  \label{fig:different_seeds} 
\end{figure} 

As \METHOD randomly selected the initial training set, it may cause the ``cold start'' problem in active learning. This means that different initial training data may lead to different initial TCR-epitope binding affinity prediction models \(M\), and in turn, may affect subsequent models and queried pairs in following iterations. To demonstrate the robustness of our method under different starting points, we conducted additional experiments based on three different random seeds and consistently observed significant amounts of annotation cost reductions (Fig.~\ref{fig:different_seeds}).

\begin{figure}[h]
  \centering 
  \includegraphics[width=\columnwidth]{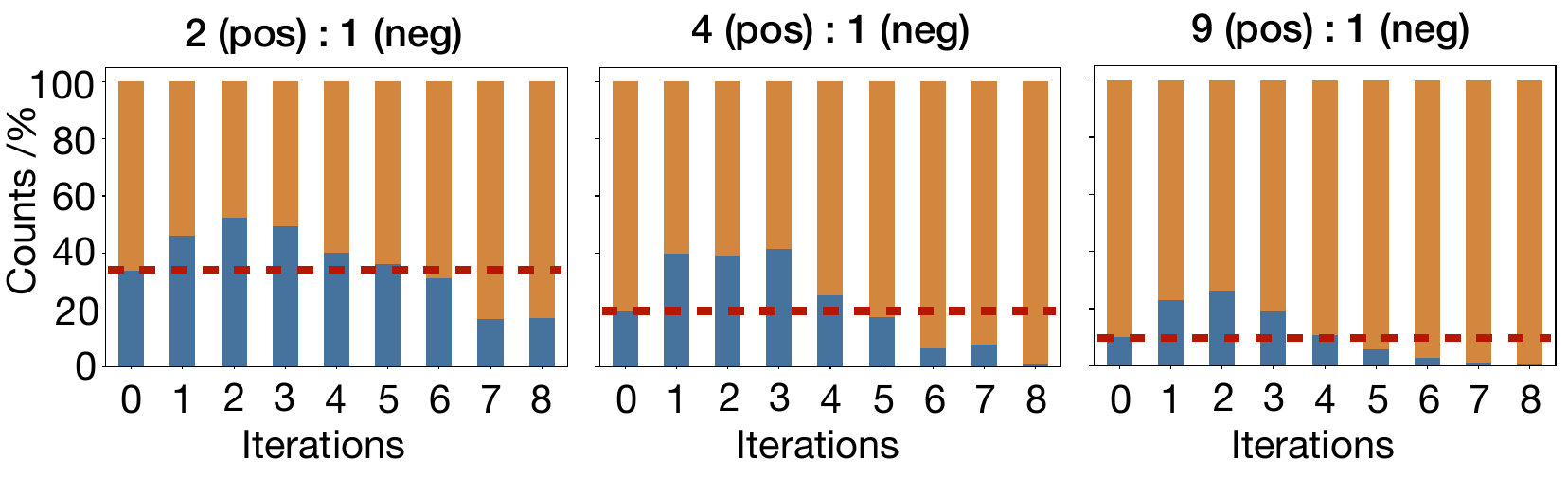} 
  \caption{Ratio of positive (colored in orange) and negative (colored in blue) pairs queried by \textit{global entropy sampling}. The initial training set had different positive-negative ratios, which are shown as red dashed lines. The left panel has a 2:1 ratio, the middle panel has a 4:1 ratio, and the right panel has a 9:1 ratio.
  }
  \label{fig:pos_neg_diff_ratio} 
\end{figure} 

One limitation of our study is that our experimental setting may not be a perfect reflection of the real-world distribution of TCR-epitope pairs. The dataset we prepared for this study has a 1:1 ratio of positive and negative pairs, while datasets obtained in real-world scenarios may differ. In a real-world scenario, researchers may focus on improving model performance for epitopes rising from a set of target diseases. Annotating a collection of TCR-epitope pairs based on a set of target epitopes may require analysis of TCR repertoire across disease and control subjects to extract TCRs that are likely to bind to the target epitopes. Consequently, there are likely to be more positive pairs and fewer negative pairs. To demonstrate how the original dataset distribution affects the query sample distributions, we ran \METHOD with global entropy sampling on a TCR-epitope dataset with different positive and negative ratios. We prepared the dataset with the ratio of positive and negative pairs as 2:1, 4:1, and 9:1, respectively. We kept the number of positive TCR-epitope pairs constant and randomly selected half, one-fourth, and one-eleventh of the negative pairs, then concatenated them together. Fig.~\ref{fig:pos_neg_diff_ratio} shows the queried sample distributions for different positive-negative ratios. While it is difficult to pinpoint the reasons, we found that the global entropy sampling query strategy attempted to query a more balanced number of positive and negative pairs than random sampling from iterations 1 to 4. This finding highlighted the potential advantage of this query strategy and we believe it deserves further investigation in future studies. We also observed that the model preferred to query positive pairs at later iterations, possibly because more positive pairs were available and negative pairs had been used up as it iterates. Additionally, the annotated TCR-epitope pairs were not specifically designed for machine learning purposes but were collected from various biological or clinical studies and may contain biases towards certain diseases.

\bibliographystyle{IEEEtran}

\onecolumn
\appendix

\makeatletter 
\renewcommand{\thefigure}{A\@arabic\c@figure}
\makeatother
\setcounter{figure}{0}

\begin{figure}[h]
  \centering 
  \includegraphics[width=.7\columnwidth]{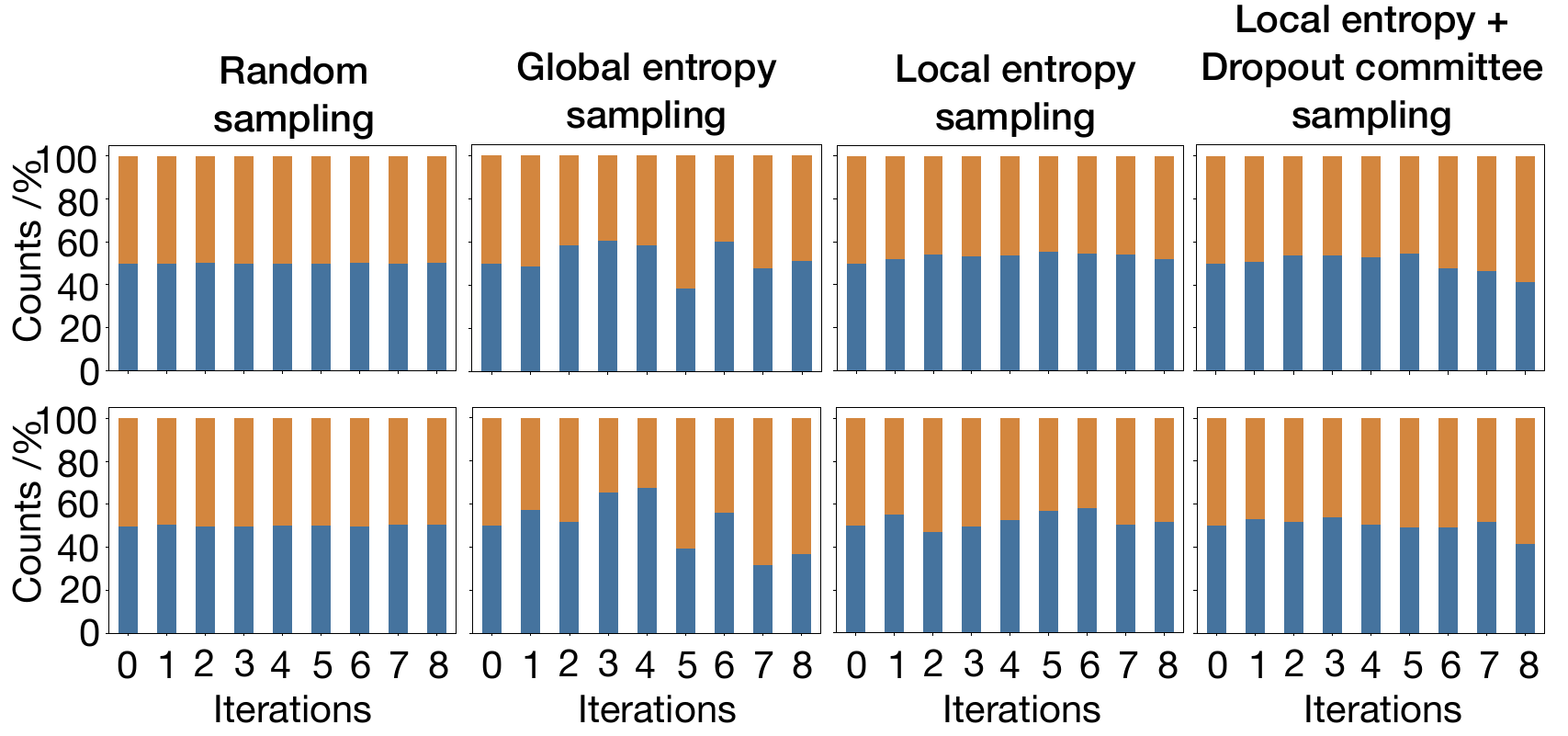} 
  \caption{Ratio of the queried positive (colored in orange) and negative (colored in blue) pairs, selected by different query strategies. The top row is For TCR split and the bottom row is for epitope split. Each column represents a query strategy.}
  \label{fig:query_distributions_all_methods} 
\end{figure} 

  \begin{figure*}[h]
  \centering 
  \includegraphics[width=.9\columnwidth]{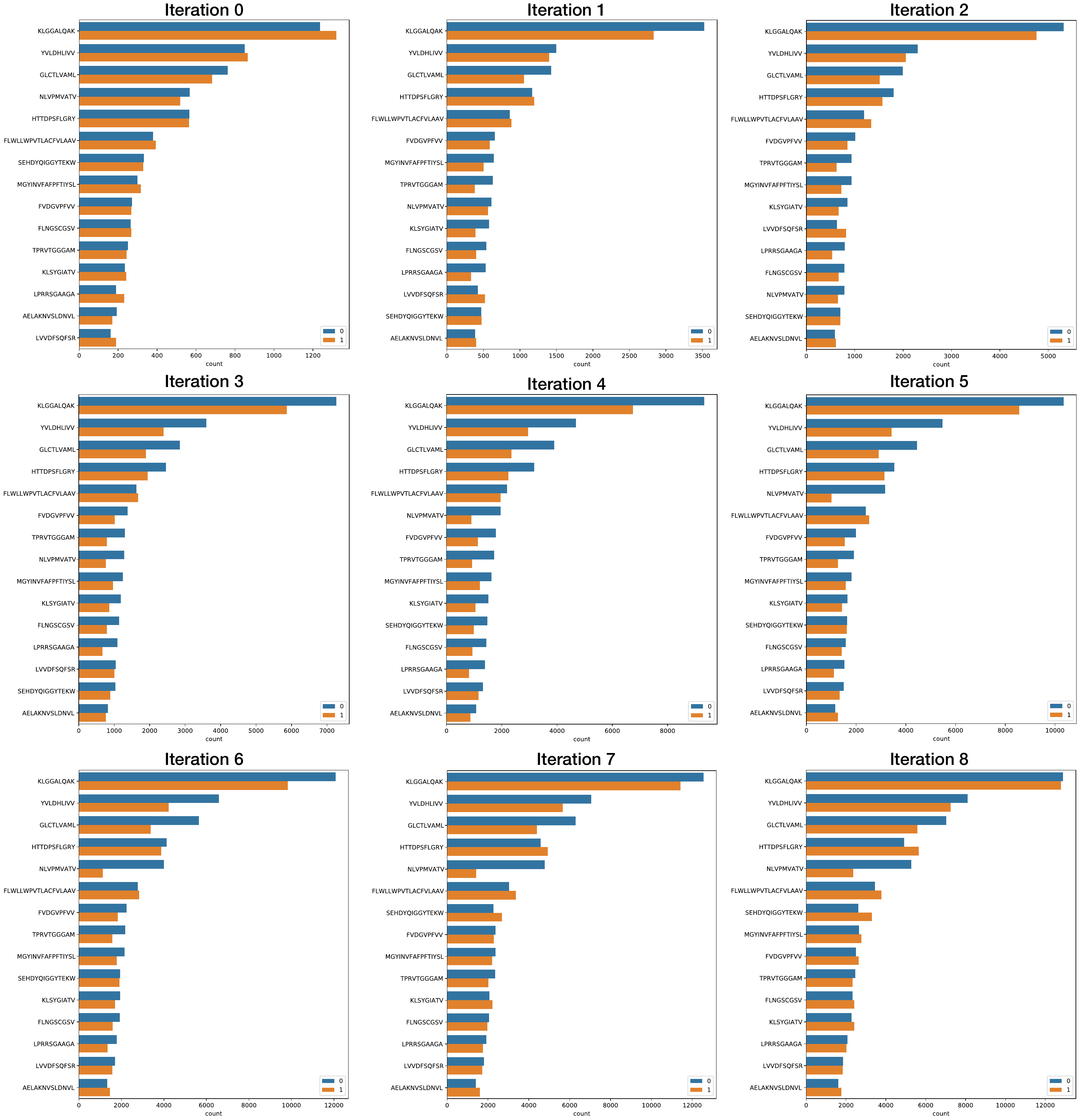} 
  \caption{Number of the queried positive (colored in orange) and negative (colored in blue) pairs, selected by \textit{global entropy sampling}. Epitope-specific numbers are measured for the top 15 frequent epitopes, and on epitope split. Values at each iteration are the cumulative number of the queried pairs [zoom in for details].
  }
  \label{fig:query_distribution_individual_epitopes} 
\end{figure*} 

\end{document}